\documentclass{article}%
\usepackage{amsmath}
\usepackage{amsfonts}
\usepackage{amssymb}
\usepackage{graphicx}
\usepackage{numinsec}
\usepackage{eurosym}
\usepackage{revsymb}
\usepackage{acronym}%
\begin{document}

\title{Construction of a General Spin Foam Model of Lorentzian BF theory and Gravity}
\author{Suresh K\ Maran}
\maketitle

\begin{abstract}
In this article we report our progress in the construction of a general
Lorentzian spin foam model based on the Gelfand-Na\u{\i}mark theory of the
representations of $SL(2,\boldsymbol{C})$ which might include both previously
proposed the $SL(2,\boldsymbol{C})/SU(2)$ based Barrett-Crane model and the
$SL(2,\boldsymbol{C})/SU(1,1).Z_{2}$ based Rovelli-Perez model. First we
construct the simplex amplitude for the BF $SL(2,\boldsymbol{C})$ model. Then
we discuss the asymptotic limit of this model. Next we discuss the
implemention of the Barrett-Crane constraints on this model. We derive an
equation that the general Lorentzian spin foam model has to satisfy. In the
appendix we give a simple derivation of the Clebsch-Gordan coefficients for
$SL(2,\boldsymbol{C})$.

\end{abstract}

\section{Introduction}

A candidate for the spin foam model\footnote{We refer to \cite{bz1} for a nice
introduction to spin foam models. We also refer to \cite{AP2} for an
up-to-date review of spin foam models and for references.} of Lorentzian
gravity was proposed by Barrett and Crane \cite{bc2}. This model was
constructed based on harmonic analysis on the homogenous space $H^{+}%
=SL(2,\boldsymbol{C})/SU(2)$ which is the upper sheet of the two-sheet
hyperboloid in four dimensional Minkowski space-time. Later Rovelli and Perez
proposed a way of deriving this model using a field theory over group
formulation \cite{R2}. But, Perez and Rovelli \cite{R4} also proposed an
alternative spin foam model based on the homogenous space $SL(2,\boldsymbol{C}%
)/U^{(-)}$ where $U^{(-)}=SU(1,1)\otimes Z_{2}$. This suggests that there must
be a model which contains both these models. We call this model the general
model while we call the Rovelli-Perez and the Barrett-Crane model the partial models.

To construct a Lorentzian spin foam model both Barrett-Crane and Rovelli-Perez
proceed by analogy with the construction of the Riemmanian spin foam model. To
explain our method for construction of the general model, let us discuss first
the construction of the spin foam model of Riemannian gravity model. The first
step is to do the path integral quantization of the $SO(4)$ BF\ theory
\cite{BF} using its discrete action to get its\ spin foam model \cite{oo1}.
Then the next step is to impose the Barrett-Crane constraints at the quantum
level. This is supposed to result in the spin foam model of the Riemannian
gravity \cite{AP2}, \cite{bc1}, \cite{bz2}, \cite{rz1}, \cite{rz2},. The third
step is to rewrite the simplex amplitude in terms of certain propagators in
the homogenous space\ $S^{3}$. To construct a Lorentzian spin foam model both
Barrett-Crane and Rovelli-Perez proceed by analogy with the third step. But
the problem with this approach is, that there are three possible subgroups for
$SL(2,\boldsymbol{C})$, which could be used for a Barrett-Crane type
construction namely $SU(2)$, $E(2)$ and $SU(1,1)$. Because of this, the
analogy cannot suggest a unique Lorentzian spin foam model. We here construct
the Lorentzian spin foam model starting from the first step using the
Gelfand-Na\u{\i}mark representation theory of $SL(2,\boldsymbol{C})$
\cite{GN1}. The assumption is that this method would yield a general spin foam
model that contain the spin foams relating to all the subgroups.We believe it
is important to do this for two reasons. First, it is good to relate a model
to an action Second, its good to have all the three subgroups $SU(2)$, $E(2)$
and $U(1,1)$ of $SL(2,\boldsymbol{C})$ play a role. Because these groups are
physically related to particles which are massive, massless and tachyonic.
Even though the third one is not physical, we donot know what role this type
of contribution would play in quantum gravity in future. If the tachyonic part
of our theory is not physical, we want our theory itself predict that, instead
of our artificially excluding it. At this point it is also not clear what role
the $E(2)$ will play in the theory. We believe the study of the asymptotic
limit \cite{Rg1}, \cite{BS}, the continuum limit and the inclusion of matter
will shed light on this. Here we restrict ourselves to the construction of
spin foam models only.

In section two we discuss how to construct the spin foam model of the
$SL(2,\boldsymbol{C})$ BF theory using the Gelfand- Na\u{\i}mark
representation theory of $SL(2,\boldsymbol{C})$ and analyze the asymptotic
limit of the BF\ spin foam model. We formally derive the equations that
describe the asymptotic limit of the theory. In section three we discuss how
to impose the Barrett-Crane constraints on the BF\ spin foam model to get a
general Lorentzian spin foam model of gravity. We present an equation that
needs to be solved to get the general Lorentzian model. In section four we
present two issues that need to be addressed in the construction of Lorentzian
spin foam models. We have tried to make this article as much self contained as possible.

\section{The Spin Foam for the SL(2,$\boldsymbol{C}$) BF Theory}

Our presentation here in the first two subsections follows that of Baez
\cite{bz1}, \cite{bz2}. The new ingredient is the use of the
Gelfand-Na\u{\i}mark represention theory of $SL(2,C)$. Advanced readers might
be able to glance through the first subsections.and the earlier part of the
second subsection.

\subsection{From the discrete action}

The Spin foam model for the $SL(2,\boldsymbol{C})$ BF\ theory action can be
derived from\ the discretized BF action by using the path integral
quantization \cite{oo1}, \cite{bz1}. Let $M$ be a simplicial manifold. Let
$g_{e}\in SL(2,\boldsymbol{C})$ be the discretized connection associated to
the edges (three-simplices) and $H_{b}=%
%TCIMACRO{\tprod _{e\supset b}}%
%BeginExpansion
{\textstyle\prod_{e\supset b}}
%EndExpansion
g_{e}$ be the holonomy around a bone (two-simplices). Then the discrete BF
action is%
\[
S_{d}=\sum_{b}tr(B_{b}H_{b}).
\]
Here $B_{b}\in sl(2,\boldsymbol{C})^{\ast}$ is the discrete analog of $B$. The
$B_{b}$ are $4\times4$ antisymmetric complex matrices. Then the\ quantum
partition function is calculated using the path integral formulation as
\footnote{While calculating the path integral only the real part of the action
is being used. Otherwise the integration with respect to the $B_{b}$ variables
no longer leads to the condition that the curvature\ (holonomy) is zero
(identity) as required by the equations of motion of the BF theory, if the
group elements are complex. The physical consequences of this procedure have
to be investigated.}%

\[
Z=\int\prod_{b}dB_{b}\exp(i\operatorname{Re}(S_{d}))\prod_{e}dg_{e}%
\]

\begin{equation}
=\int\prod_{b}\delta(H_{b})\prod_{e}dg_{e}, \label{eq.del}%
\end{equation}
where $dg_{e}$ is the invariant measure on the group $SL(2,\boldsymbol{C})$. A
summary of the Gelfand-Na\u{\i}mark representation theory of
$SL(2,\boldsymbol{C})$ and relevant references are given in appendix A.

Now consider the identity \cite{R2}
\begin{equation}
\delta(g)=\frac{1}{8\pi^{4}}\int\chi\bar{\chi}tr(T_{\chi}(g))d\chi,
\label{eq.del.exp}%
\end{equation}
where the $T_{\chi}(g)$ is the $\chi=n+i\rho$ unitary representation of
$SL(2,\boldsymbol{C})$. The Integration with respect to $d\chi$ in the above
equation is interpreted as the summation over an integer $n$ and the
integration over a real variable $\rho$. The $\left\vert \chi\right\vert
=\sqrt{n^{2}+\rho^{2}}$ is the analog of the dimension of a representation of
a compact group.

Substituting this into equation (\ref{eq.del}) we get%

\[
Z=\int\prod_{b}\left[  \int\frac{\chi_{b}\bar{\chi}_{b}}{8\pi^{4}}%
tr(T_{\chi_{b}}(\prod_{e\supset b}g_{e}))d\chi_{b}\right]  \prod_{e}dg_{e}%
\]

\[
=\int\left[  \left(  \prod_{b}\frac{\chi_{b}\bar{\chi}_{b}}{8\pi^{4}}\right)
\left(  \int\prod_{b}dg_{e}tr(\prod_{e\supset b}T_{\chi_{b}}(g_{e}))\right)
\right]  \prod_{b}d\chi_{b}%
\]

\begin{equation}
=\int\left[  \left(  \prod_{b}\frac{\chi_{b}\bar{\chi}_{b}}{8\pi^{4}}\right)
Tr\left(  \prod_{e}\int dg_{e}\bigotimes_{b\subset e}T_{\chi_{b}}%
(g_{e})\right)  \right]  \prod_{b}d\chi_{b}, \label{eq.der}%
\end{equation}
where $Tr$ denotes the required trace operations from the trace operations in
the previous line. The integrand of the quantity in the second parentheses is
the $g_{e}$ integration of the tensor product of the representation matrices
$\rho_{\chi_{b}}(g_{e})$ that are part of the holonomy around the four bones
of the edge $e$. This quantity can be rewritten as a product of the
intertwiners $i_{e}$ for $SL(2,\boldsymbol{C})$ as follows%
\begin{equation}
\int dg\bigotimes_{b\supset e}T_{\chi_{b}}(g)=\sum_{i_{e}}i_{e}\bar{\imath
}_{e}. \label{eq.5.0}%
\end{equation}
The integral on the left hand side of this equation will be referred to as the
\emph{edge integral}. The bar denotes the complex conjugation. In the right
hand side of the edge integral there is a sum of product of two intertwiners.
Each of the intertwiners corresponds to one of the sides of the edge.

Above, it is assumed that the holonomies pass through the edge in the same
direction. But usually their directions can be random. Reversing the direction
of a holonomy is equivalent to complex conjugating (the inverse of the
transpose) the representations in the edge integral. To simplify the
calculation of the edge integrals, the directions of the holonomies can be
chosen as illustrated below for a simplex. The parallel sets of arrows
indicate the direction in which the holonomies are traversed through the edges
of a simplex.
%TCIMACRO{\FRAME{dtbpF}{2.331in}{2.1048in}{0pt}{}{}{holonomy4.eps}%
%width 2.331in;  height 2.1048in;  depth 0pt;  original-width 3.9719in;
%original-height 3.5834in;  cropleft "0";  croptop "1";  cropright "1";
%cropbottom "0";  filename 'holonomy4.eps';file-properties "XNPEU";}}}%
%BeginExpansion
\begin{center}
\includegraphics[
height=2.1048in,
width=2.331in
]%
{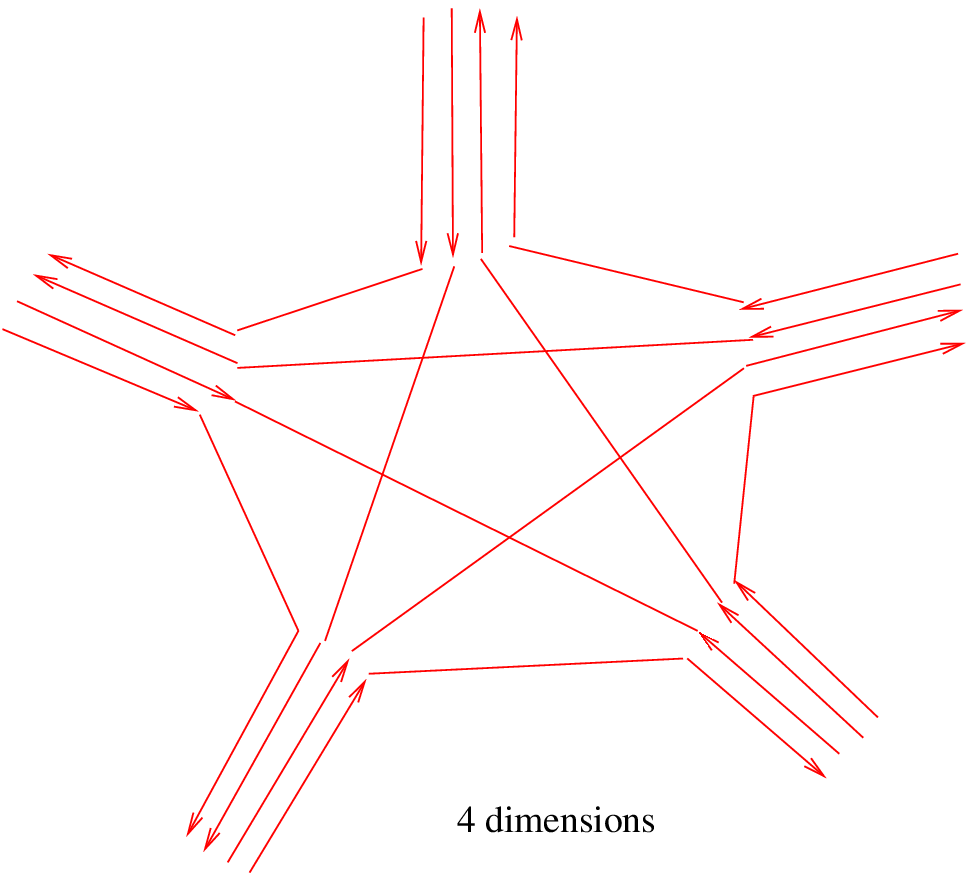}%
\end{center}
%EndExpansion

The corresponding edge integral given below is calculated as follows,%

\[
I_{4}=\int dgT_{\chi}(g)(x_{1},x_{2})T_{\lambda}(g)(y_{1},y_{2})\bar{T}_{\mu
}(g)(z_{1},z_{2})\bar{T}_{\mu}(g)(z_{1},z_{2})
\]

\[
=\int dg%
%TCIMACRO{\FRAME{itbpF}{1.1165in}{1.0334in}{0.5215in}{}{}{edge4int1.eps}%
%width 1.1165in;  height 1.0334in;  depth 0.5215in;  original-width 1.3318in;
%original-height 1.2306in;  cropleft "0";  croptop "1";  cropright "1";
%cropbottom "0";  filename 'edge4int1.eps';file-properties "XNPEU";}}}%
%BeginExpansion
\raisebox{-0.5215in}{\includegraphics[
height=1.0334in,
width=1.1165in
]%
{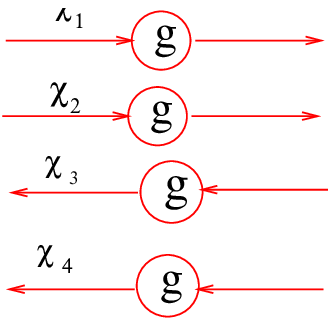}%
}%
%EndExpansion
\]

\[
=\int d^{2}\chi\int dg%
%TCIMACRO{\FRAME{itbpF}{2.0721in}{1.0672in}{0.5111in}{}{}{edge4int2.eps}%
%width 2.0721in;  height 1.0672in;  depth 0.5111in;  original-width 2.7069in;
%original-height 1.3811in;  cropleft "0";  croptop "1";  cropright "1";
%cropbottom "0";  filename 'edge4int2.eps';file-properties "XNPEU";}}}%
%BeginExpansion
\raisebox{-0.5111in}{\includegraphics[
height=1.0672in,
width=2.0721in
]%
{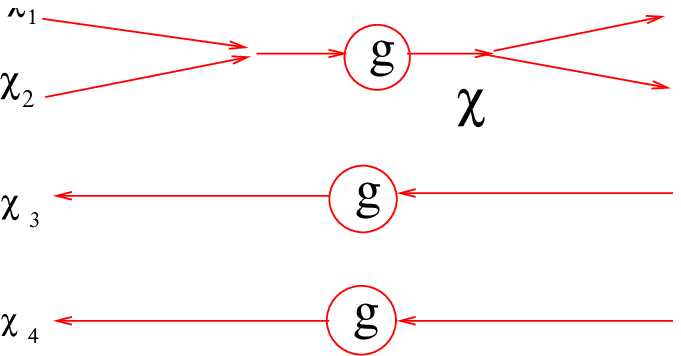}%
}%
%EndExpansion
.
\]

From appendix B we have%

\[
I_{4}=\int d^{2}\chi%
%TCIMACRO{\FRAME{itbpF}{1.9104in}{1.017in}{0.5111in}{}{}{edge4int3.eps}%
%width 1.9104in;  height 1.017in;  depth 0.5111in;  original-width 2.5668in;
%original-height 1.3534in;  cropleft "0";  croptop "1";  cropright "1";
%cropbottom "0";  filename 'edge4int3.eps';file-properties "XNPEU";}}}%
%BeginExpansion
\raisebox{-0.5111in}{\includegraphics[
height=1.017in,
width=1.9104in
]%
{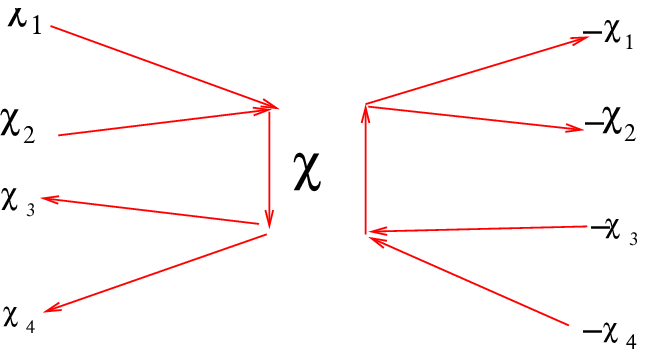}%
}%
%EndExpansion
.
\]

The nodes where the three links meet are the Clebsch-Gordan coefficients of
$SL(2,C)$. In the next section we discuss these coefficients. When this edge
integral formula is used in equation (\ref{eq.der}) and all the required trace
operations are performed, it is seen that each open link of each intertwiner
corresponding to an inner side of an edge of each simplex, only integrates
with the link of an intertwiner corresponding to an inner side of another edge
of the same simplex. The partition function $Z$ splits into a product of
terms, with each term interpreted as a quantum amplitude associated to a
simplex in the triangulation.

The quantum amplitude associated to each simplex $s$ is given below and will
be referred to as the $\{15\chi\}$ symbol,%

\[
\{15\chi\}=%
%TCIMACRO{\FRAME{itbpF}{2.111in}{2.0626in}{1.0239in}{}{}{15x.eps}%
%width 2.111in;  height 2.0626in;  depth 1.0239in;  original-width 5.5841in;
%original-height 5.4561in;  cropleft "0";  croptop "1";  cropright "1";
%cropbottom "0";  filename '15x.eps';file-properties "XNPEU";}}}%
%BeginExpansion
\raisebox{-1.0239in}{\includegraphics[
height=2.0626in,
width=2.111in
]%
{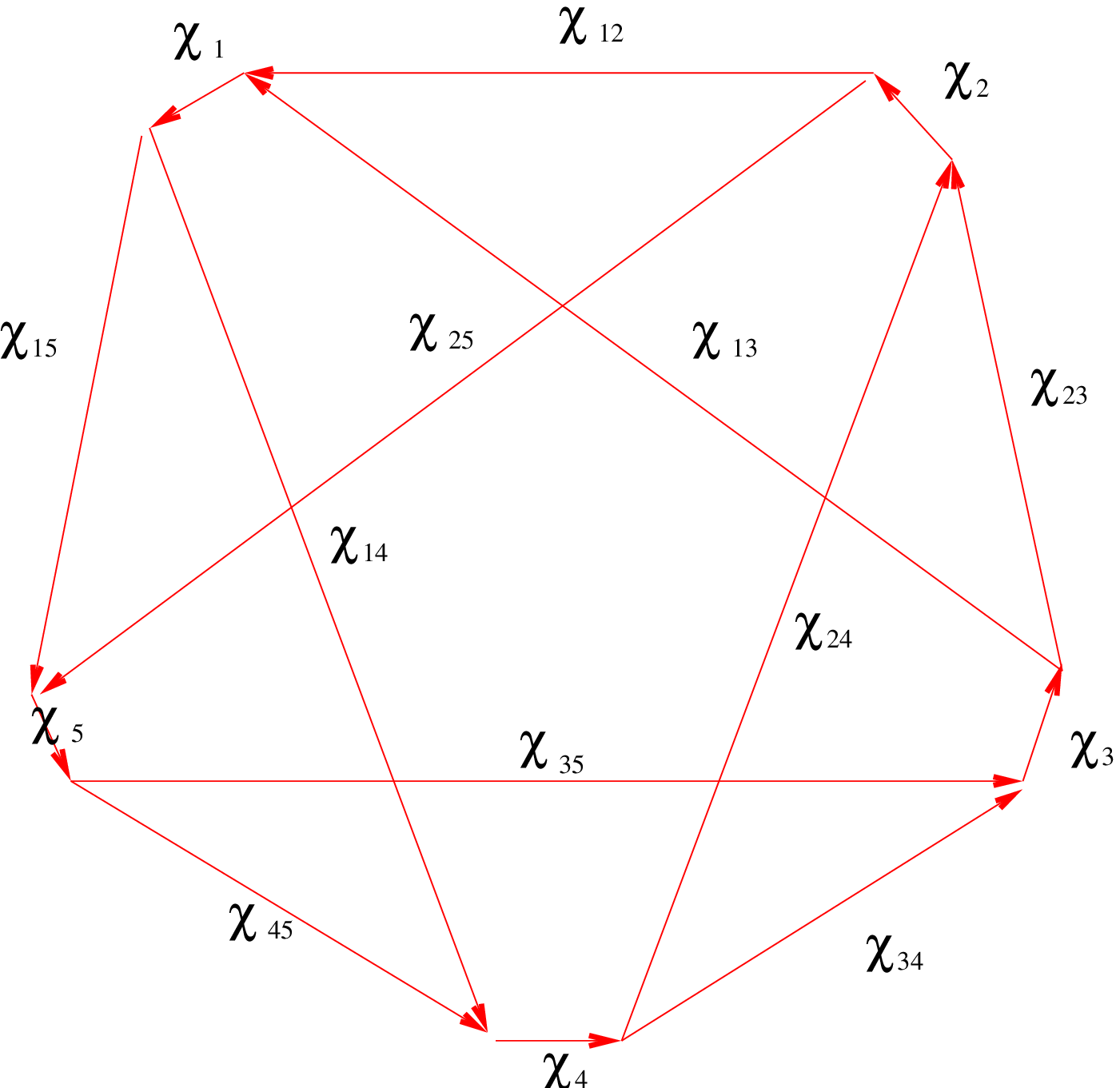}%
}%
%EndExpansion
.
\]

The$\frac{\chi_{b}\bar{\chi}_{b}}{8\pi^{4}}$ term is the quantum amplitude
associated to the bone $b$. The final partition function is
\begin{equation}
Z=\int\prod_{b}\frac{\chi_{b}\bar{\chi}_{b}}{8\pi^{4}}\prod_{s}Z(s)\prod
_{b}d\chi_{b}, \label{eq.6}%
\end{equation}
where $Z(s)$ is interpreted as the amplitude for an $n$-simplex $s$ and
$d_{\chi_{b}}$ is interpreted as the amplitude of the bone $b$, $D\chi
=\prod_{b}d\chi_{b}$ $\prod_{e}d\chi_{e}$. The integration with the measure
$d\chi$ ($\chi=m+i\rho)$ is understood as a summation of over $m$ and ordinary
integration with respect to $\rho$. \ Here $\chi_{e}$ is the internal
representation used to define the intertwiner $d\chi_{e}$. This partition
function may not be finite in general.

\subsection{Using the Quantization of the Edges.}

We believe, this version of the derivation was originally introduced for the
Riemannian Gravity by Barbieri \cite{Bar1}, and further developed by Baez,
Barrett and Crane \cite{bz2}, \cite{BB}, \cite{bc1}. These authors mainly
focussed on the group $SO(4)$. Here we apply this method that for the group
$SO(3,1)\approx SL(2,\boldsymbol{C})$.

Let $B_{i}$ $\in sl(2,\boldsymbol{C})^{\ast}$ where $i=1,2,3,4$ be the
discrete variables associated to the four bones (triangles) of an edge
(tetrahedron) of a four-simplex. Consider the equation:%

\[
B_{1}+B_{2}+B_{3}+B_{4}=0.
\]

This is the closure constraint. This is the discrete version of the equation
$dB=0$ which is one of the field equations of the continuum BF\ theory. This
equation can be quantized as follows. To each bone $i$ of the edge associate a
Hilbert space $H_{i}$ (which is the linear space that carries the unitary
representation of $SL(2,\boldsymbol{C})$) on which the quantum generators
$\hat{B}_{i}$ (which are simply the unitary representations of generators of
$SL(2,\boldsymbol{C})$) act. Then the previous equation can be promoted to a
quantum constraint,%
\[
\left(  \hat{B}_{1}+\hat{B}_{2}+\hat{B}_{3}+\hat{B}_{4}\right)  \psi=0,
\]
where $\psi\in H_{1}\otimes H_{2}\otimes H_{3}\otimes H_{4}$. The solutions of
this equation are the intertwiners $i_{e}$,
\begin{equation}
\psi=%
%TCIMACRO{\FRAME{itbpF}{1.5833in}{1.3154in}{0.6182in}{}{}{4x.eps}%
%width 1.5833in;  height 1.3154in;  depth 0.6182in;  original-width 2.2627in;
%original-height 1.8742in;  cropleft "0";  croptop "1";  cropright "1";
%cropbottom "0";  filename '4X.eps';file-properties "XNPEU";}}}%
%BeginExpansion
\raisebox{-0.6182in}{\includegraphics[
height=1.3154in,
width=1.5833in
]%
{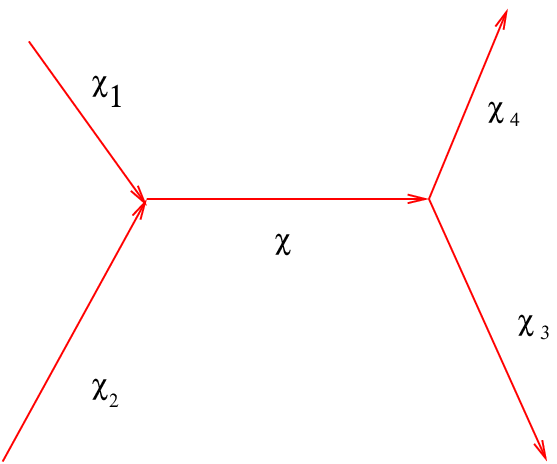}%
}%
%EndExpansion
, \label{eq.4j}%
\end{equation}
where it is assumed that each edge $i$ is associated a $D_{\chi_{i}}$ the
linear space of unitary representation $\chi_{i}$.

Each solution of the constraint is a quantum state that depends firstly on the
five complex numbers $\chi_{1},\chi_{2},\chi_{3},\chi_{4}$ and $\chi$ each of
which labels a unitary representation of $SL(2,\boldsymbol{C})$ and secondly
on a 2-2 split of the four bones in the edges. In the above diagram a 12-34
split is being used.

In the graph, in equation, the three arrowed links intersecting at the nodes
are the Clebsch-Gordan coefficients of $SL(2,\boldsymbol{C})$. Changing the
direction of the arrows is equivalent to complex conjugating the associated representation.

The Clebsch-Gordan coefficients were originally derived by Na\u{\i}mark
\cite{N1}. A quick way to calculate them is shown in appendix B. The
Clebsch-Gordan coefficients are\ explicitly%

\[%
%TCIMACRO{\FRAME{itbpF}{1.4062in}{1.42in}{0.7022in}{}{}{3x.eps}%
%width 1.4062in;  height 1.42in;  depth 0.7022in;  original-width 4.2307in;
%original-height 4.2756in;  cropleft "0";  croptop "1";  cropright "1";
%cropbottom "0";  filename '3X.eps';file-properties "XNPEU";}}}%
%BeginExpansion
\raisebox{-0.7022in}{\includegraphics[
height=1.42in,
width=1.4062in
]%
{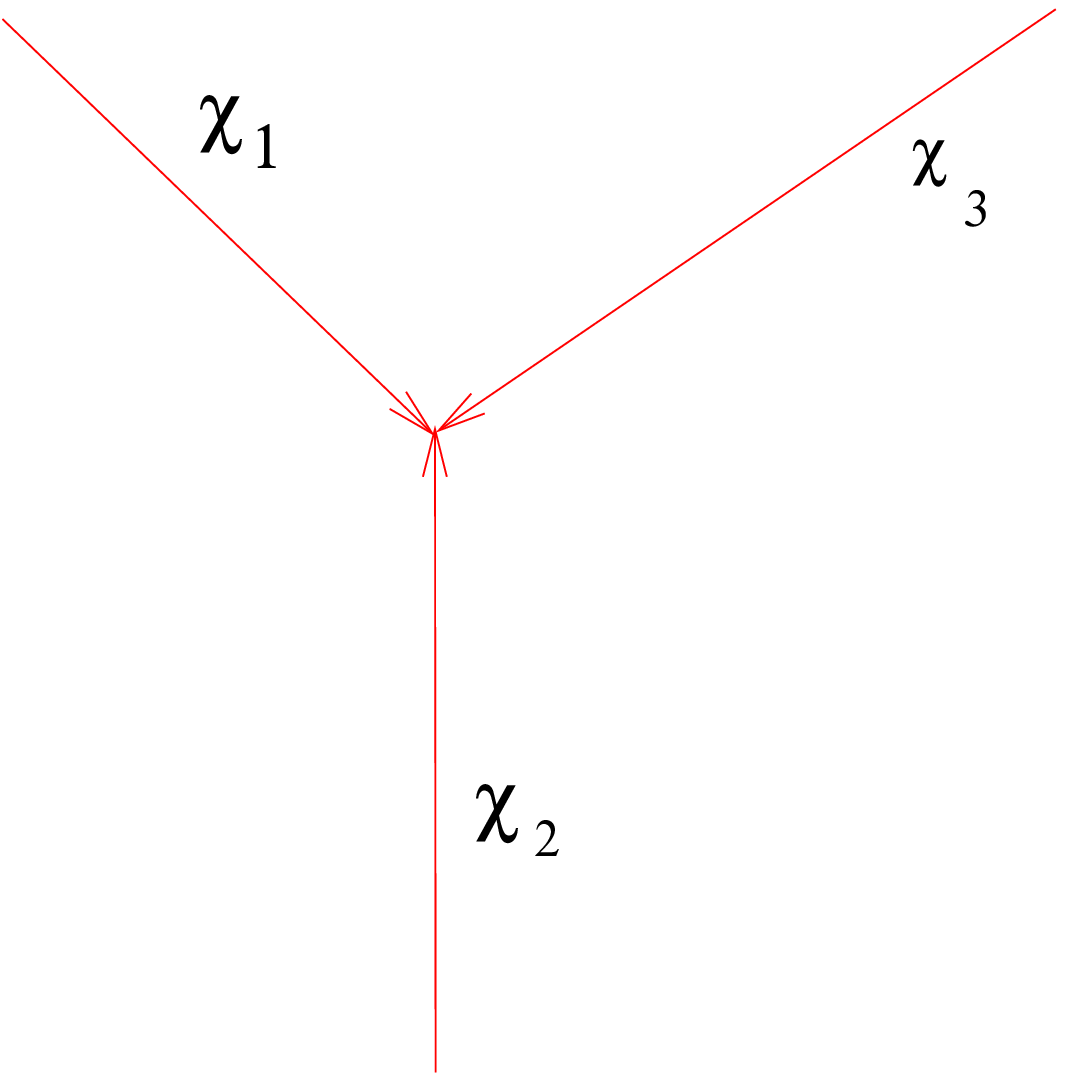}%
}%
%EndExpansion
=C(\chi_{1},\chi_{2},\chi_{3},z_{1},z_{2},z_{3})
\]

\begin{align*}
&  =\left(  z_{1}-z_{2}\right)  ^{\frac{-\chi_{1}-\chi_{2}+\chi_{3}}{2}%
-\frac{1}{2}}\left(  z_{2}-z_{3}\right)  ^{\frac{\chi_{1}-\chi_{2}-\chi_{3}%
}{2}-\frac{1}{2}}\left(  z_{3}-z_{1}\right)  ^{\frac{-\chi_{1}+\chi_{2}%
-\chi_{3}}{2}-\frac{1}{2}}\\
&  \left(  \bar{z}_{1}-\bar{z}_{2}\right)  ^{-\frac{-\bar{\chi}_{1}-\bar{\chi
}_{2}+\bar{\chi}_{3}}{2}-\frac{1}{2}}\left(  \bar{z}_{2}-\bar{z}_{3}\right)
^{-\frac{\bar{\chi}_{1}-\bar{\chi}_{2}-\bar{\chi}_{3}}{2}-\frac{1}{2}}\left(
\bar{z}_{3}-\bar{z}_{1}\right)  ^{-\frac{-\bar{\chi}_{1}+\bar{\chi}_{2}%
-\bar{\chi}_{3}}{2}-\frac{1}{2}}.
\end{align*}

Let $\frac{n_{1}}{2},\frac{n_{2}}{2}$ and$\frac{n_{3}}{2}$ are the real parts
of $\chi_{1},\chi_{2}$ and $\chi_{2}$. The above result is to be replaced by
zero if the sum $n_{1}+n_{2}+n_{3}$ is not even (please see appendix B).
Usually in the $SU(2)\ $spin networks the internal link stands for a summation
over the azimuthal quantum numbers. Here it is replaced by integration over
the $z$'s using the Riemann measure on the complex plane $dz\wedge d\bar{z}$.

Now a quantum amplitude can be calculated for any closed simplicial
$3$-surface made of tetrahedra in the same way as it is calculated for a
closed $2$-surface using the Clebsch-Gordan coefficients of $SU(2)$ \cite{RP}.
The 4-simplex amplitude `the $\{15\chi\}$ symbol' is calculated by evaluating
the spin network associated to its boundary.

This simplex amplitude is precisely analogous to the one calculated in the
previous section. \newpage

\subsection{Asymptotic limit of the BF simplex amplitude}

The $\{15\chi\}$ symbol is given by%

\[
\{15\chi\}=%
%TCIMACRO{\FRAME{itbpF}{2.1318in}{2.0833in}{1.0239in}{}{}{15x.eps}%
%width 2.1318in;  height 2.0833in;  depth 1.0239in;  original-width 5.5841in;
%original-height 5.4561in;  cropleft "0";  croptop "1";  cropright "1";
%cropbottom "0";  filename '15x.eps';file-properties "XNPEU";}}}%
%BeginExpansion
\raisebox{-1.0239in}{\includegraphics[
height=2.0833in,
width=2.1318in
]%
{15x.eps}%
}%
%EndExpansion
.
\]

In this diagram $\chi_{i}$ is the unitary representation associated to the
internal link of each edge, $\chi_{ij}$ is the representation associated to
the bone which is the intersection of $i$'th edge and $j$'th edge. The $z_{i}$
and $z_{ij}$ are similarly defined. The asymptotic limit is the limit of the
$\{15\chi\}$ symbol when the $\chi$'s are scaled by a real number $\lambda$
which is sent to infinity \cite{Rg1}, \cite{BS}.

The simplex amplitude is of the form
\begin{equation}
\{15\chi\}=\int a(z_{i},z_{ij})\exp(-i\phi(z_{i},z_{ij},\chi_{i},\chi_{ij}))%
%TCIMACRO{\tprod _{i}}%
%BeginExpansion
{\textstyle\prod_{i}}
%EndExpansion
dz_{i}%
%TCIMACRO{\tprod _{i<j}}%
%BeginExpansion
{\textstyle\prod_{i<j}}
%EndExpansion
dz_{ij}, \label{eq.15x}%
\end{equation}
where $a(z_{i},z_{ij})$ are factors in $\{15\chi\}$ that do not depend the
$\chi$'s in the definition of the Clebsch-Gordan coefficients, $\phi
(z_{i},z_{ij},\chi_{i},\chi_{ij})$ is $i$ times the natural log of the product
of the remaining factors. The $\phi(z_{i},z_{ij},\chi_{i},\chi_{ij})$ is a
real function which is first order in the $\chi$'s and $\ln(z_{i}-z_{jk})$ and
takes the form
\[
\sum\chi_{l}\ln(z_{i}-z_{jk})+c.c,
\]
where summation is over specific choices of the $\chi_{l}\ln(z_{i}-z_{jk})$
terms which can be explicitly calculated from the definition of $\{15\chi\}$.
The asymptotic limit is the limiting function of the following when
$\lambda\longrightarrow\infty$
\[
\{15\chi\}_{\lambda}=\int a(z_{i},z_{ij})\exp(-i\lambda\phi(z_{i},z_{ij}%
,\chi_{i},\chi_{ij}))%
%TCIMACRO{\tprod _{i}}%
%BeginExpansion
{\textstyle\prod_{i}}
%EndExpansion
dz_{i}%
%TCIMACRO{\tprod _{i<j}}%
%BeginExpansion
{\textstyle\prod_{i<j}}
%EndExpansion
dz_{ij}.
\]

The limit can be calculated using the geometric asymptotic formula \cite{GS},%
\[
\int a(y)e^{i\lambda\phi(y)}dy=\left(  \frac{2\pi}{k}\right)  ^{\frac{n}{2}%
}\sum_{y|d\phi(y)=0}e^{\frac{\pi isgnH(y)}{4}}\frac{e^{ik\phi(y)}a(y)}%
{\sqrt{\det\left\vert H(y)\right\vert }}+O(k^{-n/2-1}),
\]

where $y$ is an real $n$-tuple and $H(y)$ is the Hessian of $\phi$ at $y$.
This implies the asymptotics are controlled by the extremums of $\phi(y)$. In
equation (\ref{eq.15x}) $n$ is $30$.

Since $\phi(z_{i},z_{ij},\chi_{i},\chi_{ij})$ is real in equation
(\ref{eq.15x}), its extrema are determined just by the equations%

\[
\frac{\partial\phi}{\partial z_{i}}=0,\frac{\partial\phi}{\partial z_{ij}%
}=0,\forall i,j.
\]

These equations are of the form,
\begin{equation}
\sum\frac{\chi_{l}}{(z_{i}-z_{jk})}=0. \label{eq.ass}%
\end{equation}

Further work is need to be done to solve and investigate solutions of these
equations which we leave as a open problem.\ We believe this work can help us
understand the semiclassical limit of the related gravity and also the
relationship between gravity and topological field theories in four dimensions.

\section{The Spin Foam Model of SL(2,$\boldsymbol{C}$)\ Gravity.}

In the previous section a spin foam for the Lorentzian BF theory was derived.
To reduce this theory to that of gravity further constraints, called the
Barrett-Crane constraints given below have to be imposed at the quantum level
on the edges \cite{bc1}, \cite{bz2}, \cite{rz1}, \cite{rz2}, \cite{AP2}%

\[
B_{i}\wedge B_{j}=0,\forall i,j.
\]
The Barrett-Crane constraints are basically the discretized Plebanski
constraints \cite{Pleb}.

In the BF theory each simplex is an atom of a topology. Imposing the above
constraints reduces the atom of a topology to an atom of a geometry. The
solution for these equations in the case of Riemannian gravity was proposed by
Barrett-Crane, partially proved by Barbieri \cite{Bar2} and completely proved
by Reisenberger \cite{rz2} .

To sovle Barrett-Crane constraints for Lorentzian gravity, consider the
following ansatz for the solution of the Barrett-Crane constraints%
\begin{equation}
\psi_{\chi_{1}\chi_{2}\chi_{3}\chi_{4}}=\int d\chi f(\chi)%
%TCIMACRO{\FRAME{itbpF}{1.3396in}{1.1122in}{0.5422in}{}{}{4x.eps}%
%width 1.3396in;  height 1.1122in;  depth 0.5422in;  original-width 2.2061in;
%original-height 1.8273in;  cropleft "0";  croptop "1";  cropright "1";
%cropbottom "0";  filename '4X.eps';file-properties "XNPEU";}}}%
%BeginExpansion
\raisebox{-0.5422in}{\includegraphics[
height=1.1122in,
width=1.3396in
]%
{4X.eps}%
}%
%EndExpansion
. \label{eq.bc.ver}%
\end{equation}
Now for any $B$ $\in$ $sl(2,\boldsymbol{C})^{\ast}$, the quantum operator
$\hat{B}$ can be used to construct the two Casimir operators of
$SL(2,\boldsymbol{C})$. The mathematics required has been explained by Ruhl in
Ref.\cite{RL}. The components of $\hat{B}$ are the rotation and the boost
generators $\hat{J}_{i}$ and $\hat{F}_{i}$ ( $i=1,2,3$ ). The \^{B} is given by,%

\[
\hat{B}=\left[
\begin{array}
[c]{cccc}%
0 & F_{1} & F_{2} & F_{3}\\
-F_{1} & 0 & J_{3} & -J_{2}\\
-F_{2} & -J_{3} & 0 & J_{1}\\
-F_{3} & J_{2} & -J_{1} & 0
\end{array}
\right]  .
\]

Define $F_{\pm}=F_{1}\pm F_{2},$ $H_{\pm}=H_{1}\pm H_{2}$. These operators are
given explicitly in appendix A. If $A_{ab},B_{cd}$ are four dimensional
antisymmetric tensors define $A\wedge B=\varepsilon^{abcd}$ $A_{ab}B_{cd}$.
Then the Casimir operators are the following \cite{RL}:%

\begin{equation}
I_{1}=\hat{B}\wedge\ast\hat{B}=F_{+}F_{-}+F_{-}F_{+}+2F_{3}^{2}-H_{+}%
H_{-}-H_{-}H_{+}-2H_{3}^{2},
\end{equation}

\begin{equation}
I_{2}=\hat{B}\wedge\hat{B}=H_{+}F_{-}+H_{-}F_{+}+F_{+}H_{-}+F_{-}H_{+}%
+4H_{3}F_{3}, \label{eq.i2}%
\end{equation}

where $H_{\pm}=H_{1}\pm iH_{2}$ and $F_{\pm}=F_{1}\pm iF_{2}.$

These two operators are proportional to the identity operator in their actions
on the functions $\in D_{\chi}$ with the eigenvalues $\frac{1}{2}\left(
\rho^{2}-m^{2}-2\right)  $ and $\rho m$.

So the classical Barrett-Crane constraint at the quantum level for $i=j$,
$\hat{B}_{i}\wedge\hat{B}_{i}\psi=0,$ simply states that either $\rho_{i}$ or
$m_{i}$ is zero. This was first observed by Barrett-Crane \cite{bc1}. So each
$i$, these constraints can be solved by setting $\rho_{i}$ or $m_{i}$ to be
zero. The Barrett-Crane model of the Lorentzian quantum gravity corresponds to
the case in which $m_{i}=$ $0$ for all edges. In the Rovelli-Perez model
\cite{R4} a specific linear superposition of both of these cases $\rho_{i}=0$
and $m_{i}=0$, which is sufficient to do harmonic analysis on the single sheet
hyperboloid of four dimensional Minkowski space-time \cite{GMV} was used.

Hence forth we assume that either $\rho_{i}$ or $m_{i}$ is zero.

Consider the graph in equation (\ref{eq.bc.ver}). The $\hat{B}_{1}\wedge
\hat{B}_{2}\psi=0$ ( and $\hat{B}_{3}\wedge\hat{B}_{4}\psi=0$ ) can be solved
as follows. Consider
\[
\left(  \hat{B}_{1}+\hat{B}_{2}\right)  \wedge\left(  \hat{B}_{1}+\hat{B}%
_{2}\right)  =\hat{B}_{1}\wedge\hat{B}_{1}+\hat{B}_{2}\wedge\hat{B}_{2}%
+2\hat{B}_{1}\wedge\hat{B}_{2}.
\]
From this it can be inferred that
\[
\left(  \hat{B}_{1}+\hat{B}_{2}\right)  \wedge\left(  \hat{B}_{1}+\hat{B}%
_{2}\right)  \psi=0\Longrightarrow\hat{B}_{1}\wedge\hat{B}_{2}\psi=0.
\]
Due to the invariance of the three vertex in equation (\ref{eq.bc.ver}), this
suggests setting $\rho$ or $m$ $=$ $0$ where $\chi=$ $m+i\rho$ is the
representation of the internal link.

Imposing $\hat{B}_{1}\wedge\hat{B}_{3}\psi=0$ ( or one of $\hat{B}_{1}%
\wedge\hat{B}_{4}\psi=0$ , $\hat{B}_{2}\wedge\hat{B}_{3}\psi=0$ and $\hat
{B}_{2}\wedge\hat{B}_{4}\psi=0)$ is a more difficult case. It can be evaluated
explicitly in terms of the generators using the identity,%

\begin{equation}
\hat{B}_{1}\wedge\hat{B}_{3}=\frac{\hat{B}_{1}\wedge\hat{B}_{1}+\hat{B}%
_{3}\wedge\hat{B}_{3}-\left(  \hat{B}_{1}+\hat{B}_{3}\right)  \wedge\left(
\hat{B}_{1}+\hat{B}_{3}\right)  }{2}.
\end{equation}
To evaluate the right hand side the results given in equation (\ref{eq.i2})
can be used. Then the expression for the generators as operators on the
representation space $D_{\chi}$ derived by Ruhl is substituted. The resulting
calculation is cumbersome, but the result is simple and is given by%

\begin{equation}
\hat{B}_{1}\wedge\hat{B}_{3}=\left(  \bar{z}_{1}-\bar{z}_{3}\right)  ^{2}%
\frac{\partial}{\partial\bar{z}_{1}}\frac{\partial}{\partial\bar{z}_{3}%
}-\left(  z_{1}-z_{3}\right)  ^{2}\frac{\partial}{\partial z_{1}}%
\frac{\partial}{\partial z_{3}}+i\left(  \frac{m_{1}\rho_{3}+m_{3}\rho_{1}}%
{2}-\rho_{1}-\rho_{3}\right)  . \label{eq.sol}%
\end{equation}

One must solve the equation $\left(  \hat{B}_{1}\wedge\hat{B}_{3}\right)
\psi=0$ (and other similar cross constraints) for $\psi$. This will enable us
to construct the general Lorentzian model.

\section{Two Issues}

\subsection{{\protect\small A problem with the representation theory}}

One of the problems that we encountered in our work is with the expansion of
$\delta(g)$. Having a proper expansion for $\delta(g)$ is equivalent to having
a proper harmonic analysis for the functions on the group $SL(2,\boldsymbol{C}%
)$. One can derive the following expansion for $\delta(g)$ using the
Gelfand-Na\u{\i}mark representation theory%

\[
\delta(g)=\frac{1}{8\pi^{4}}\int\chi\bar{\chi}tr(\rho_{\chi}(g))d\chi.
\]

This expansion was first used by Rovelli and Perez \cite{R2}. In the
right-hand side the $tr(\rho_{\chi}(g))$ is purely a function of the
eigenvalues of $g$ (please see appendix A). Because of this right-hand-side of
the above equation is peaked at $g$'s for which the eigen-values are $1$. But
the eigen-values of $g$ is $1$ not only at the identity but also when it is
strictly upper or lower triangular. So this means that the right-hand-side
does not have the proper expansion of $\delta(g)$. This expansion for
$\delta(g)$ has been used here and also in the Lorentzian spin foam
derivations by Rovelli and Perez \cite{R2}, \cite{R4}. We believe that this
problem has to be fixed or clarified.

\subsection{{\protect\small Rigorous derivation of the imposition of the
constraints.}}

In section three we proceeded to derive a spin foam model of Lorentzian
quantum gravity by imposing the quantum Barrett-Crane constraints on the BF
spin foam model. But both in Riemannian and Lorentzian quantum gravity, this
way of imposing of the Barrett-Crane constraints is not yet been rigorously
derived using the path integral quantization of any discrete action. There
have been various proposals \cite{LF1}, \cite{Reis1} for doing this
calculation but they have not yet been fully implemented. Because of this
issue we believe the amplitudes of the lower dimensional ($<4$) simplices are
yet to be fixed (assumming we got the four-simplex amplitudes correct atleast
to a certain level). Please read \cite{AP3} for more ideas about this.

\section{Conclusion}

Here a construction of the general Lorentzian spin foam model has been
explained. To finish the model one would have to solve equation (\ref{eq.sol}%
). This will enable us to construct a model which is possibly richer in
physical aspects compared to the previously known models. Also we formally
derived the equations that describe the asymptotic limit of $SL(2,C)$ BF
theory. These equations should be solvable because of their simplicity.
Investigating the solutions of these will help us understand the asymptotic
limit of the general Lorentzian model. Also they can shed more light on the
relation between the topological field theory and gravity in four
dimensions.\textit{ But before getting oneself into investigating all these,
we believe it is important to address the two issues mentioned in the previous
section.}

\section{Acknowledgements}

I am grateful to George Sparling for discussions, encouragement and guidance
in developing this article and in learning the material. I also thank Allen
Janis for discussions and support. I thank the Department of Physics and
Astronomy, University of Pittsburgh for encouragements and financial support.

\appendix{}

\section{The representation theory of SL(2,$\boldsymbol{C}$)}

The Representation theory of $SL(2,\boldsymbol{C})$ was developed by Gelfand
and Na\u{\i}mark \cite{GN1}, \cite{GMV}. This was further studied and
developed by Ruhl \cite{RL}. We recommend \cite{GMV} to the beginners. The
unitary irreducible representations of $SL(2,\boldsymbol{C})$ are infinite
dimensional. Each unitary representation of\ $SL(2,\boldsymbol{C})\ $can be
described by an appropriate action of the group on functions of single complex
variable $\phi(z)$ and is labelled by a complex number $\chi=\frac{n}%
{2}+i\frac{\rho}{2}$, where $n$ is an integer and $\rho$ is a real number. We
denote the linear space on which the representation acts as $D_{\chi}$.
Henceforth we denote $\chi$ as a pair $(\chi_{1},\chi_{2})$ where $\chi
_{1}=-\bar{\chi}_{2}=\frac{n+i\rho}{2}$. The readers might refer to the
references for a fuller descrition of $D_{\chi}$ and all other related
mathematical constructs.

Let $g$ is an element of $SL(2,\boldsymbol{C})$ given by%

\[
g=\left[
\begin{array}
[c]{cc}%
\alpha & \beta\\
\gamma & \delta
\end{array}
\right]  ,
\]
where $\alpha$,$\beta$,$\gamma$ and $\delta$ are complex numbers such that
$\alpha\delta-\beta\delta=1$.

Then the $\chi$ representation can be described by action of an unitary
operator $T_{\chi}(g)$ on functions $\phi(z$ $)$ of a complex variable $z$ as
given below.%

\[
T_{\chi}(g)\phi(z)=(\beta z_{1}+\delta)^{\chi_{1}-1}(\beta^{\ast}z_{1}^{\ast
}+\delta^{\ast})^{\chi_{2}-1}\phi(\frac{\alpha z+\gamma}{\beta z+\delta}).
\]

This action on $\phi(z)$ is unitary under the inner product defined by%

\[
\left(  \phi(z),\eta(z)\right)  =\frac{i}{2}\int d^{2}z\bar{\phi}%
(z)\eta(z)d^{2}z,
\]
where $d^{2}z=dz\wedge d\bar{z}$ .

The above equation can also be written as,%

\[
T_{\chi}(g)\phi(z_{1})=\int T_{\chi}(g)(z_{1},z_{2})\phi(z_{2})d^{2}z_{2}.
\]

The $T_{\chi}(g)(z_{1},z_{2})$ is defined as%

\begin{equation}
T_{\chi}(g)(z_{1},z_{2})=(\beta z_{1}+\delta)^{\chi-1}(\beta^{\ast}z_{1}%
^{\ast}+\delta^{\ast})^{-\bar{\chi}-1}\delta(z_{2}-g(z_{1})), \label{eq.rep}%
\end{equation}

where $g(z_{1})=\frac{\alpha z_{1}+\gamma}{\beta z_{1}+\delta}$

The Kernel $T_{\chi}(g)(z_{1},z_{2})$ is the analog of the matrix
representation of finite dimensional unitary representations of compact
groups. We will denote this diagrammatically by$%
%TCIMACRO{\FRAME{itbpF}{1.0205in}{0.2101in}{0.0813in}{}{}{edge1int.eps}%
%{\special{ language "Scientific Word";  type "GRAPHIC";  display "USEDEF";
%valid_file "F";  width 1.0205in;  height 0.2101in;  depth 0.0813in;
%original-width 1.3509in;  original-height 0.2883in;  cropleft "0";
%filename 'edge1int.eps';file-properties "XNPEU";}}}%
%BeginExpansion
\raisebox{-0.0813in}{\includegraphics[
height=0.2101in,
width=1.0205in
]%
{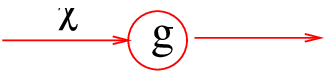}%
}%
%EndExpansion
$.

An infinitesimal group element $a$ of $SL(2,\boldsymbol{C})$ can be
parametrized by six real real numbers $\varepsilon_{k}$ and $\eta_{k}$ as
shown below \cite{RL}%

\[
a\approx I+\frac{i}{2}\sum_{k=1}^{3}(\varepsilon_{k}\sigma_{k}+\eta_{k}%
i\sigma_{k}),
\]

where the $\sigma_{k}$ are the Pauli matrices. The corresponding six
generators of the $\chi$ representation are $H_{k}$ and $F_{k}$ . The $H_{k}$
correspond to rotations and the $F_{k}$ correspond to boosts. These are
differential operators on $\phi(z)$ which are defined as%

\[
F_{\pm}=F_{1}\pm iF_{2},\qquad H_{\pm}=H_{1}\pm iH_{2},
\]

\[
H_{+}=-z^{2}\frac{\partial}{\partial z}-\frac{\partial}{\partial\bar{z}}%
+(\chi_{1}-1)z,\qquad F_{-}=i\frac{\partial}{\partial z}-i\bar{z}^{2}%
\frac{\partial}{\partial\bar{z}}+i(\chi_{2}-1)\bar{z},
\]%
\[
H_{-}=\bar{z}^{2}\frac{\partial}{\partial\bar{z}}+\frac{\partial}{\partial
z}-(\chi_{2}-1)\bar{z},\qquad F_{+}=-iz^{2}\frac{\partial}{\partial z}%
+i\frac{\partial}{\partial\bar{z}}+i(\chi_{1}-1)z,
\]%
\[
H_{3}=z\frac{\partial}{\partial z}-\bar{z}\frac{\partial}{\partial\bar{z}%
}+\frac{1}{2}(\chi_{2}-\chi_{1}),\qquad F_{3}=iz\frac{\partial}{\partial
z}+i\bar{z}\frac{\partial}{\partial\bar{z}}-\frac{i}{2}(\chi_{2}+\chi_{1}-2).
\]

A measure on the group $g$ is given by
\[
dg=\left(  \frac{i}{2}\right)  ^{3}\frac{d^{2}\beta d^{2}\gamma d^{2}\delta
}{\left\vert \delta\right\vert ^{2}}=\left(  \frac{i}{2}\right)  ^{3}%
\frac{d^{2}\alpha d^{2}\beta d^{2}\gamma}{\left\vert \alpha\right\vert ^{2}}.
\]
This measure is invariant under left translation, right translation and
inversion in $SL(2,\boldsymbol{C})$.

The measure can be also written as follows

\qquad\qquad\qquad\qquad%
\[
dg=\left(  \frac{i}{2}\right)  ^{3}\delta(\alpha\beta-\gamma\delta
-1)d^{2}\alpha d^{2}\beta d^{2}\gamma d^{2}\delta,
\]

where we allow$\ \alpha,\beta,\gamma$ and $\delta$ to vary freely while doing integration.

The Fourier transform theory on $SL(2,\boldsymbol{C})$ was developed in
\cite{GMV}. If $f(g)$ is a square integrable function on the group, it has a
group Fourier transform defined by%

\[
F(\chi)=\int f(g)T_{\chi}(g)dg.
\]

The associated inverse Fourier transform is%

\[
f(g)=\frac{1}{8\pi^{4}}\int F(\chi)T_{\chi}(g^{-1})\chi_{1}\chi_{2}d\chi.
\]

From this it can inferred that%

\[
\delta(g)=\frac{1}{8\pi^{4}}\int tr\left[  T_{\chi}(g)\right]  \chi_{1}%
\chi_{2}d\chi.
\]

The above equation was first used by Rovelli and Perez \cite{R2}, \cite{R4} to
derive the spin foam models of Lorentzian gravity from the formalism of field
theory over group.

\section{Derivation of the 3D edge integral.}

We need to evaluate the following edge integral which is needed to determine
the four dimensional edge integral and also the Clebsch-Gordan Coefficients of
$SL(2,\boldsymbol{C})$.%
\begin{align*}
I_{3}  &  =\int%
%TCIMACRO{\FRAME{itbpF}{1.1451in}{0.9287in}{0.4124in}{}{}{edge3int1.eps}%
%width 1.1451in;  height 0.9287in;  depth 0.4124in;  original-width 1.3509in;
%original-height 1.0901in;  cropleft "0";  croptop "1";  cropright "1";
%cropbottom "0";  filename 'edge3int1.eps';file-properties "XNPEU";}}}%
%BeginExpansion
\raisebox{-0.4124in}{\includegraphics[
height=0.9287in,
width=1.1451in
]%
{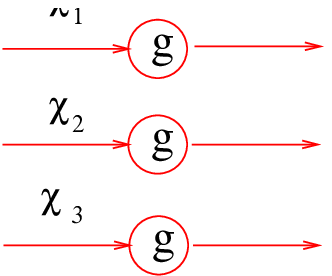}%
}%
%EndExpansion
dg\\
&  =\int T_{\chi}(g)(x_{1},x_{2})T_{\lambda}(g)(y_{1},y_{2})T_{\mu}%
(g)(z_{1},z_{2})dg
\end{align*}

$\qquad$%
\begin{align*}
&  =\int(\beta x_{1}+\delta)^{\chi_{1}-1}(\beta^{\ast}x_{1}^{\ast}%
+\delta^{\ast})^{\chi_{2}-1}\delta(x_{2}-\frac{\gamma+x_{1}\alpha}%
{\delta+x_{1}\beta})\\
&  (\beta y_{1}+\delta)^{\lambda_{1}-1}(\beta^{\ast}y_{1}^{\ast}+\delta^{\ast
})^{\lambda_{2}-1}\delta(y_{2}-\frac{\gamma+y_{1}\alpha}{\delta+y_{1}\beta})\\
&  (\beta z_{1}+\delta)^{\mu_{1}-1}(\beta^{\ast}z_{1}^{\ast}+\delta^{\ast
})^{\mu_{2}-1}\delta(z_{2}-\frac{\gamma+z_{1}\alpha}{\delta+z_{1}\beta})dg
\end{align*}
$\qquad\qquad\qquad$%
\begin{align*}
&  =\int(\beta x_{1}+\delta)^{\chi_{1}}(\beta^{\ast}x_{1}^{\ast}+\delta^{\ast
})^{\chi_{2}}\delta(\left(  \delta+x_{1}\beta\right)  x_{2}-\gamma-x_{1}%
\alpha)\\
&  (\beta z_{1}+\delta)^{\mu_{1}}(\beta^{\ast}z_{1}^{\ast}+\delta^{\ast}%
)^{\mu_{2}}\delta(\left(  \delta+z_{1}\beta\right)  z_{2}-\gamma-z_{1}%
\alpha)\\
&  (\beta y_{1}+\delta)^{\lambda_{1}}(\beta^{\ast}y_{1}^{\ast}+\delta^{\ast
})^{\lambda_{2}}\delta(\left(  \delta+y_{1}\beta\right)  y_{2}-\gamma
-y_{1}\alpha)\delta\left(  \gamma\beta-\alpha\delta+1\right) \\
&  d^{2}\alpha d^{2}\beta d^{2}\gamma d^{2}\delta.
\end{align*}

Let us define the variables $a=\frac{\alpha}{\gamma},b=\frac{\beta}{\gamma}$
and $c=\frac{\delta}{\gamma}.$

To eliminate the deltas we need to solve the following linear equations for
$a,$ $b$ and $c:$%

\begin{align*}
\left(  c+x_{1}b\right)  x_{2}-x_{1}a  &  =1,\\
\left(  c+y_{1}b\right)  y_{2}-y_{1}a  &  =1,\\
\left(  c+z_{1}b\right)  z_{2}-z_{1}a  &  =1.
\end{align*}

The determinant of the above system is%

\[
D=x_{1}x_{2}y_{1}z_{2}-x_{1}x_{2}y_{2}z_{1}-x_{1}y_{1}y_{2}z_{2}+x_{2}%
y_{1}y_{2}z_{1}+x_{1}y_{2}z_{1}z_{2}-x_{2}y_{1}z_{1}z_{2}.
\]

The solution of the linear system is given by%

\begin{align*}
a  &  =\frac{x_{1}x_{2}y_{2}-x_{1}x_{2}z_{2}-x_{2}y_{1}y_{2}+x_{2}z_{1}%
z_{2}+y_{1}y_{2}z_{2}-y_{2}z_{1}z_{2}}{x_{1}x_{2}y_{1}z_{2}-x_{1}x_{2}%
y_{2}z_{1}-x_{1}y_{1}y_{2}z_{2}+x_{2}y_{1}y_{2}z_{1}+x_{1}y_{2}z_{1}%
z_{2}-x_{2}y_{1}z_{1}z_{2}},\\
b  &  =\frac{x_{2}y_{1}-x_{1}y_{2}+x_{1}z_{2}-x_{2}z_{1}-y_{1}z_{2}+y_{2}%
z_{1}}{x_{1}x_{2}y_{2}z_{1}-x_{1}x_{2}y_{1}z_{2}+x_{1}y_{1}y_{2}z_{2}%
-x_{2}y_{1}y_{2}z_{1}-x_{1}y_{2}z_{1}z_{2}+x_{2}y_{1}z_{1}z_{2}},\\
c  &  =\frac{x_{1}x_{2}z_{1}-x_{1}x_{2}y_{1}+x_{1}y_{1}y_{2}-x_{1}z_{1}%
z_{2}-y_{1}y_{2}z_{1}+y_{1}z_{1}z_{2}}{x_{1}x_{2}y_{2}z_{1}-x_{1}x_{2}%
y_{1}z_{2}+x_{1}y_{1}y_{2}z_{2}-x_{2}y_{1}y_{2}z_{1}-x_{1}y_{2}z_{1}%
z_{2}+x_{2}y_{1}z_{1}z_{2}}.
\end{align*}

Let us define the variables,%
\begin{align*}
A  &  =\left(  x_{1}-z_{1}\right)  \left(  y_{1}-x_{1}\right)  \left(
z_{2}-y_{2}\right)  ,\\
B  &  =\left(  x_{2}-z_{2}\right)  \left(  z_{1}-y_{1}\right)  \left(
y_{1}-x_{1}\right)  ,\\
C  &  =\left(  y_{2}-x_{2}\right)  \left(  y_{1}-z_{1}\right)  \left(
z_{1}-x_{1}\right)  ,\\
t  &  =\frac{\gamma}{D},\\
E  &  =\left(  z_{2}-x_{2}\right)  \left(  z_{2}-y_{2}\right)  \left(
x_{2}-y_{2}\right)  \left(  z_{1}-y_{1}\right)  \left(  z_{1}-x_{1}\right)
\left(  y_{1}-x_{1}\right)  .
\end{align*}
Then using the expressions for $a,b$ and $c$ we can derive the following
results,%
\begin{align*}
c+x_{1}b  &  =D^{-1}A,\\
c+y_{1}b  &  =D^{-1}B,\\
c+z_{1}b  &  =D^{-1}C,\\
b-ac  &  =D^{-2}E.
\end{align*}

Using the above results and definitions, $I\ $can simplified as,%

\[
I_{3}=A^{\chi_{1}}\bar{A}^{\chi_{2}}B^{\lambda_{1}}\bar{B}^{\lambda_{2}}%
C^{\mu_{1}}\bar{C}^{\mu_{2}}\int\delta\left(  t^{2}E+1\right)  t^{\chi
_{1}+\lambda_{1}+\mu_{1}}\bar{t}^{\chi_{1}+\lambda_{1}+\mu_{1}}d^{2}t.
\]

Let $\frac{n_{1}}{2}$,$\frac{n_{2}}{2}$ and$\frac{n_{3}}{2}$ be the imaginary
parts of $\chi_{1},\lambda_{1}$ and $\mu_{1}$. We can solve for $t$ from
$t^{2}E+1=0$ as $\pm E^{-\frac{1}{2}}$. Substituting this we find $I_{3}\ $is
not zero only if $n_{1}+n_{2}+n_{3}$ is even. Hence forth let us assume that
$n_{1}+n_{2}+n_{3}$ is even.

After the final integration we get
\[
I_{3}=2A^{\chi_{1}}\bar{A}^{\chi_{2}}B^{\lambda_{1}}\bar{B}^{\lambda_{2}%
}C^{\mu_{1}}\bar{C}^{\mu_{2}}\left\vert E\right\vert ^{-1}E^{-\frac{\chi
_{1}+\lambda_{1}+\mu_{1}}{2}}\bar{E}^{-\frac{\chi_{1}+\lambda_{1}+\mu_{1}}{2}%
}.
\]
From this $I_{3}$ can been explicitly expressed as%

\begin{align*}
I_{3}  &  =C(\mu_{1},\chi_{1},\lambda_{1},x_{1},y_{1},z_{1})C(-\mu_{2}%
,-\chi_{2},-\lambda_{2},x_{2},y_{2},z_{2})\\
&  =C(\mu_{1},\chi_{1},\lambda_{1},x_{1},y_{1},z_{1})\bar{C}(\mu_{2},\chi
_{2},\lambda_{2},x_{2},y_{2},z_{2})\\
&  =%
%TCIMACRO{\FRAME{itbpF}{1.6631in}{1.1149in}{0.5145in}{}{}{edge3int2.eps}%
%{\special{ language "Scientific Word";  type "GRAPHIC";  display "USEDEF";
%valid_file "F";  width 1.6631in;  height 1.1149in;  depth 0.5145in;
%original-width 1.8378in;  original-height 1.1061in;  cropleft "0";
%filename 'edge3int2.eps';file-properties "XNPEU";}}}%
%BeginExpansion
\raisebox{-0.5145in}{\includegraphics[
height=1.1149in,
width=1.6631in
]%
{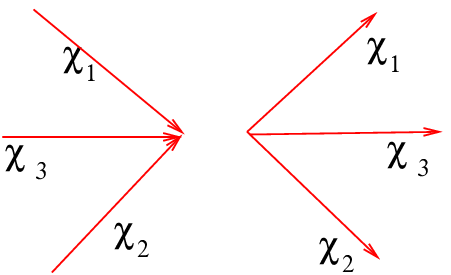}%
}%
%EndExpansion
,
\end{align*}

where $C$ is the Clebsch-Gordan coefficient defined by
\begin{align*}
C(\mu,\chi,\lambda,x,y,z)  &  =%
%TCIMACRO{\FRAME{itbpF}{1.3439in}{1.3578in}{0.5111in}{}{}{3x.eps}%
%width 1.3439in;  height 1.3578in;  depth 0.5111in;  original-width 4.2307in;
%original-height 4.2756in;  cropleft "0";  croptop "1";  cropright "1";
%cropbottom "0";  filename '3X.eps';file-properties "XNPEU";}}}%
%BeginExpansion
\raisebox{-0.5111in}{\includegraphics[
height=1.3578in,
width=1.3439in
]%
{3X.eps}%
}%
%EndExpansion
\\
&  =\left(  x-y\right)  ^{^{\frac{-\chi-\lambda+\mu}{2}-\frac{1}{2}}}\left(
z-x\right)  ^{\frac{-\chi+\lambda-\mu}{2}-\frac{1}{2}}\left(  y-z\right)
^{\frac{\chi-\lambda-\mu}{2}-\frac{1}{2}}\\
&  \overline{\left(  x-y\right)  }^{-\frac{-\bar{\chi}-\bar{\lambda}+\bar{\mu
}}{2}-\frac{1}{2}}\overline{\left(  z-x\right)  }^{-\frac{-\bar{\chi}%
+\bar{\lambda}-\bar{\mu}}{2}-\frac{1}{2}}\overline{\left(  y-z\right)
}^{-\frac{\bar{\chi}-\bar{\lambda}-\bar{\mu}}{2}-\frac{1}{2}}.
\end{align*}
We recite again that last result is to be replaced by zero if $n_{1}%
+n_{2}+n_{3}$ is not even.

\end{document}